\documentclass[conference]{IEEEtran}
\usepackage{amsmath,graphicx, multirow, dblfloatfix}
\usepackage{flushend}
\usepackage[hyphens]{url}
\usepackage{soul,color}

\ifCLASSINFOpdf
\else
\fi

\begin{document}

\title{Stacked Convolutional and Recurrent Neural Networks for Bird Audio Detection}
\author{\IEEEauthorblockN{Sharath Adavanne, Konstantinos Drossos, Emre~\c{C}ak{\i}r, Tuomas Virtanen}
\IEEEauthorblockA{Department of Signal Processing, Tampere University of Technology, Finland\\
Email: firstname.lastname@tut.fi}}
\maketitle

\begin{abstract}
This paper studies the detection of bird calls in audio segments using stacked convolutional and recurrent neural networks. Data augmentation by blocks mixing and domain adaptation using a novel method of test mixing are proposed and evaluated in regard to making the method robust to unseen data. The contributions of two kinds of acoustic features (dominant frequency and log mel-band energy) and their combinations are studied in the context of bird audio detection. Our best achieved AUC measure on five cross-validations of the development data is 95.5\% and 88.1\% on the unseen evaluation data.
\end{abstract}

\IEEEpeerreviewmaketitle

\section{Introduction}
\label{sec:intro}
Bird audio detection (BAD) refers to identifying the presence or absence of a bird call/tweet in a given audio recording. This task acts as a preliminary step in the automatic monitoring of biodiversity~\cite{Marques2012,Furnas2014}. After identifying the presence of bird call activity, a species-based classifier can recognize the bird call more accurately~\cite{Graciarena2011,Stowell2014}. In this regard, the bird audio detection challenge~\cite{bad2016} was organized with an objective to create algorithms that are robust and scalable to work on real life bio-acoustics monitoring projects without any manual intervention. The challenge provided annotated and non-annotated bird call recordings. The former is utilized as the training dataset and the latter are recordings from a completely different geographical location and employed as the test dataset. This geographical mismatch imposes a further difficulty to the problem since any proposed method should be context independent. 

The bio-diversity changes widely across geographical locations. For example, bird species in one location are not the same in the other. Different locations also mean different acoustic environments leading to a variety of sound sources specific to the respective soundscapes. Furthermore, each of these bird species has unique calls, resulting in a wide variety of bird calls. Labeling such a wide variety of calls into one class weakens the classifier and can result in misclassification of similar sounding non-bird sounds. The problem is further intensified in the dataset used because each of the bird calls has been recorded with different devices that add their own system noise. A bird audio detection method which can work across such a wide range of species and environments is termed as a generic method.

To our knowledge, there has not been any publication specific to detection of bird calls in audio. Bird audio detection has been used as a submodule in the bird species classification task~\cite{Graciarena2011,Stowell2014}. In the context of manual annotation of audio for very large biodiversity surveys~\cite{Furnas2014}, using a binary bird audio detector helps filter a number of negative instances, thereby improving the efficiency.

In this paper, we propose the employment of methods of sound event detection (SED) and their adaptation to the specific problem of detecting bird calls, approaching the BAD as a SED problem. In the case of general SED, the state of the art results have been reported in~\cite{emre_TASLP2016} using convolutional recurrent neural networks (CRNNs). The CRNN architecture exploits the combined modeling capacities of a convolutional neural network (CNN), a recurrent neural network (RNN), and a fully connected (FC) layer. CRNN architectures have also been proposed in automatic speech recognition~\cite{sainath2015} and music classification~\cite{Choi2016}. In~\cite{Adavanne2017}, these CRNN's were extended to accommodate multiple feature classes and the feature maps from CNNs were processed using a bidirectional RNN. This architecture was called the convolutional bidirectional recurrent neural network (CBRNN). We use this CBRNN for identifying the presence of bird call in the audio.

In particular, for the BAD task, we propose to use the CBRNN and train it with regularization methods like dropout and early stopping to reduce the over-fitting to training data. This makes it generic and it performs equally well on unseen data from different recording conditions. Data augmentation method of blocks mixing and a novel domain adaptation method of test mixing are proposed and analyzed with respect to making the classifier robust to new data. Two features (log mel-band energy and dominant frequency) and their combination are analyzed in the context of the BAD task.

The rest of the paper is organized as follows. The proposed method involving the extraction of acoustic features representing the harmonic and non-harmonic content of the audio are presented in Section~\ref{sec:feat}. The state of the art network for SED task and its configuration for the BAD is explained and presented in Section~\ref{sec:crnn}. Data augmentation and domain adaptation techniques are studied for generalizing the BAD methods in Section~\ref{ssec:dataaugm}. The evaluation and results are reported and discussed in section \ref{sec:eval}.

\section{Method}
The input to the proposed method is an audio signal of length 10 seconds. Acoustic features, namely log mel-band energy and dominant frequency, are extracted from this audio in frames of 40~ms. This amounts to 500 frames in total for 10 seconds audio. The stacked neural network reads in the 500 frames of features and maps them to the presence or absence of a bird call. This stacked neural network is built by stacking layers of CNN, RNN and FC followed by a single node output layer producing outputs in the range of [0, 1]. The output zero marks the presence and one marks the absence of the bird call. The details of the feature extraction and the stacked neural network are described below.

\subsection{Feature extraction}\label{sec:feat}
%Bird call is a natural phenomenon like human speech and is used by them to communicate with birds of the same species. 
In this paper, we experiment with two kinds of features and analyze their contributions. Just like human speech and singing, bird calls can have harmonic, non-harmonic, broadband, and noisy structure~\cite{Nowicki1997}.  We propose to model the overall content of the audio using the log mel-band energy feature ($mbe$). $mbe$ has also been shown to be effective in the general SED tasks~\cite{emre_TASLP2016}.

The harmonic content in the audio is proposed to be represented using three local dominant frequencies and their respective magnitudes ($dom$-$freq$) in each frame. $dom$-$freq$ has been used as a perceptual feature in SED tasks~\cite{Adavanne2017} and has provided considerable improvement when used along with $mbe$.

Both the features were extracted from frames of 40 ms length with 50\% overlap using a Hamming window. The three $dom$-$freq$'s were extracted in the range of 500-8000 Hz. By choosing a minimum frequency of 500 Hz, we get rid of most environmental ambience and human speech related fundamental frequecies. The extraction was done on thresholded parabolically-interpolated STFT~\cite{jos} using the librosa implementation~\cite{librosa}. The log mel-band energy was calculated for 40 mel-bands in 0-22050 Hz range.

\begin{figure}
  \centering
  \includegraphics[width=0.98\columnwidth]{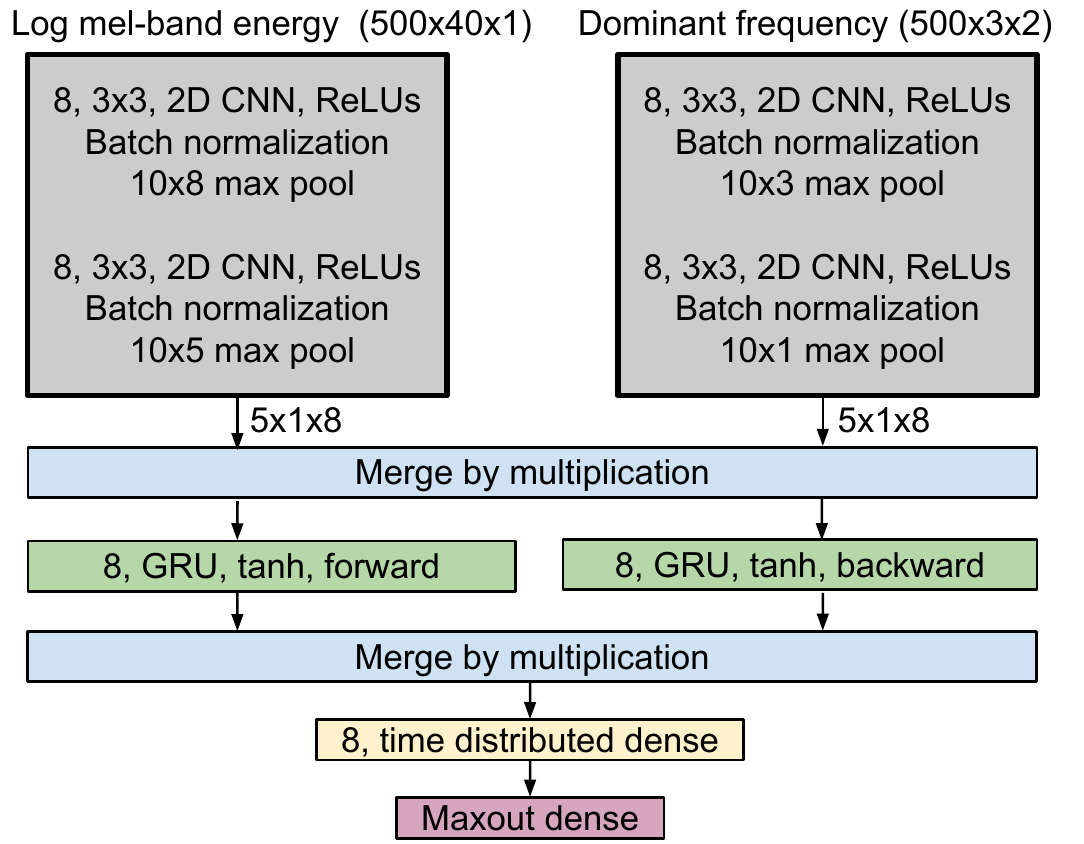}
  \caption{Stacked convolutional and bi-directional recurrent neural networks (CBRNN) architecture for bird audio detection using multiple feature classes}
  \label{fig:crnn}
\end{figure}

\subsection{Proposed neural network}
\label{sec:crnn}

Each of the feature classes, $mbe$ and $dom$-$freq$, is handled separately in the first layer of the CBRNN. $T$ = 500 frames of $40$ $mbe$ from mono channel audio are stacked into a volume of $T\times40\times1$. While the three frequencies and their amplitudes of $dom$-$freq$ are layered into a volume of $T\times3\times2$. Separate CNNs are employed to learn local shift-invariant features from each of these volumes as shown in Figure \ref{fig:crnn}. A max pooling operation is performed after every CNN layer in time and frequency axes reducing the final dimension of both the feature classes to  $5\times1\times N$, where $N$ is the number of filters in the last CNN layer. We use a receptive field of $3\times3$ for all CNNs. The feature maps from the individual CNNs are merged using an elementwise multiplication operation and fed to bi-directional gated recurrent unit (GRU) layers followed by fully-connected time distributed dense layers. The output layer consists of a maxout dense layer \cite{Goodfellow2013} with sigmoid activation function.

Batch normalization~\cite{batchNorm} was employed for all the CNN layers. The CBRNN was trained for a maximum of 500 epochs, using Adam optimizer (with the parameters proposed in the original paper)~\cite{adamKeras}, and mean squared error objective. In order to reduce overfitting of the model, early stopping was used to stop training if the area under curve (AUC) measured (Section~\ref{ssec:data}) on the validation data did not improve for 50 epochs. Dropout \cite{Dropout} was employed as a regularizer to make the model generic and avoid overfitting to the training data. The neural network architecture was implemented using Keras~\cite{chollet2015keras} and Theano backend~\cite{theano}.

\subsection{Data augmentation and domain adaptation}
\label{ssec:dataaugm}
In order to increase the generalization and robustness of our classifier, we perform data augmentation using the blocks mixing implementation of \cite{giam2016}. The features of every training file are mixed with the features of another random training file. The mixing of $dom$-$freq$ of two files is done by concatenation, this extends the feature dimension to $T\times6\times2$. In the case of $mbe$, the maximum value for each time and frequency bin is used, thereby keeping the input dimension unchanged. The network is trained with the augmented data along with the original features. This doubles the training data size. The label for the augmented data is set to be absent only if the bird call is absent in both the random files, otherwise the label is set to present. 

In the BAD challenge, since the evaluation data is from an entirely different location, the performance of the classifier on it may be poor. In order to teach the classifier what it can expect, we propose a novel approach for domain adaptation called test mixing. We perform this by exposing the network to test data by selectively mixing it with training data. Since we do not have the labels of test data, we cannot mix every training recording with a random test recording. Hence, we perform the mixing only on training recordings where bird call is present (positive label). This way no matter what content the test recording has, the training label will remain positive after mixing. Ideally, we can mix every training recording with each of the test recordings, but we limit ourselves to mixing each training recording with just one test recording. Thereby we double the amount of training data for the positive class. In future, a similar augmentation method will have to be devised for the negative cases, so that the classifier is equally exposed to test data ambiance for both the classes. 

We submitted another method~\cite{Cakir2017} which came second in the BAD challenge. The proposed method differs from~\cite{Cakir2017} in terms of using a harmonic specific feature ($dom$-$freq$), a network supporting multiple feature classes, max pooling operation in time axis and processing the feature map from CNNs using bi-directional GRU. Additionally, we also propose using data augmentation and domain adaptation to generalize our method.

\section{Evaluation} 
\label{sec:eval}

\subsection{Datasets and metrics}
\label{ssec:data}
The bird audio detection challenge \cite{bad2016} provided a development and an evaluation set. These data came from three separate datasets: i) field (freefield1010), ii) crowd-sourced (warblr), and iii) remote monitored (chernobyl). The development set comprised of freefield1010 and warblr only. The evaluation (challenge) set comprised of data unseen in development, predominantly coming from the chernobyl dataset.

Recordings in both the sets were 10 seconds long, single channel, and sampled at 44.1 kHz. The labels for the development set were binary, i.e. bird call(s) present or absent. The development set consisted of 15690 recordings in total and was distributed as presented in Table \ref{Table:1}. The evaluation set consisted of 8620 audio recordings.

\begin{table}[h]
\caption{Bird audio detection challenge~\cite{bad2016} development set statistics}
\label{Table:1}
\centering
\begin{tabular}{l|ll}
\multirow{2}{*}{Dataset} & \multicolumn{2}{c}{Bird call} \\\cline{2-3}
                         & present        & absent       \\\hline
freefield1010            & 5755           & 1935         \\
warblr                   & 1955           & 6045         \\\hline
Total                    & 7710           & 7980        \\
\end{tabular}
% \vspace{10pt}
\end{table}

From the development set, we randomly generated five cross-validation (CV) splits of 60\% training, 20\% validation, and 20\% testing such that each split had equal distribution of classes. All development set results in future are the average performance on this five-fold CV split.

For the challenge submission, the CBRNN is trained on three CV splits of 80\% training and 20\% validation of development set, with equal distribution of classes in each split. For each of the CV splits, the trained CBRNN is evaluated on the unseen test set, and the average of the three outputs is submitted as the final result.

The output of the BAD method is evaluated from the receiver operating characteristic curve (ROC) using the AUC measurement \cite{auc}.

\begin{table*}
% \fontsize{9}{12}\selectfont
\centering
% \vspace{15pt}
\caption{Area under curve scores for validation split of development dataset and unseen test data. The best test scores for each feature and dropout combination is highlighted.}
\label{Table:2}

\begin{tabular}{c|c|cc|cc|cc}
\multicolumn{2}{c}{} & \multicolumn{2}{|c|}{No data augmentation} & \multicolumn{2}{c|}{Blocks mixing}& \multicolumn{2}{c}{Test mixing} \\\hline

Feature                    & Dropout & \begin{tabular}[c]{@{}c@{}}Validation\\ Score\end{tabular} & \begin{tabular}[c]{@{}c@{}}Test \\ Score\end{tabular} & \begin{tabular}[c]{@{}c@{}}Validation\\ Score\end{tabular} & \begin{tabular}[c]{@{}c@{}}Test \\ Score\end{tabular} & \begin{tabular}[c]{@{}c@{}}Validation\\ Score\end{tabular} & \begin{tabular}[c]{@{}c@{}}Test \\ Score\end{tabular} \\\hline

\multirow{3}{*}{$mbe$ + $dom$-$freq$} & 0.25 & 95.1& 85.0 & 94.9& 83.2 & 94.6 & \bf86.5 \\
 & 0.5 & 94.7& 85.6 & 95.5& 83.4 & 94.6& \bf87.4 \\
 & 0.75& 94.8& 83.7 & 95.2& 85.3 & 94.7& \bf86.2 \\\hline

\multirow{3}{*}{$mbe$} & 0.25& 95.2& 87.2 & 95.0& 84.8 & 94.8& \bf88.1 \\
 & 0.5 & 95.3& 85.1 & 95.2& 86.5 & 94.9& \bf87.6 \\
 & 0.75& 95.3& 87.0 & 95.4& 86.1 & 94.7& \bf87.8                                              
\end{tabular} 
% \vspace{-10pt}
\end{table*}

\subsection{Evaluation procedure}
For the estimation of the hyper-parameters of the CBRNN, we experimented with one to four layers each of CNN, RNN, and FC. The number of units for each of these layers were varied in the set of $\{4, 8, 16, 32, 64, 128\}$. The same dropout rate was used for all layers and varied in the set of $\{0.25, 0.50, 0.75\}$. The parameters were decided based on the best AUC score on five CVs of the development set, using the $mbe$ and $dom$-$freq$ features. The best configuration with least number of weights had two layers of CNNs with eight filters each, one RNN layer with eight units and an FC with eight units. Figure \ref{fig:crnn} shows the configuration and the feature map dimensions of the neural network. This configuration had only 2,600 weights. In terms of AUC score, configurations of CBRNN having up to 500,000 weights did not show any significant improvement over using 2,600 weights.

The best CBRNN configuration was seen to generalize well with a dropout of 0.75 and was seen to overfit for 0.25 and 0.50. The overfitting was observed from the training and validation AUC score plot with respect to training epochs. On employing early stopping, we control this overfitting at different drop out rates and achieve a comparable AUC on the development set. 

Similar hyper-parameter experiments were done for the $mbe$ and $dom$-$freq$ features individually, and the same CBRNN configuration was seen to be one of the top performers on the development set with over 95\% AUC for $mbe$ and around 87\% for $dom$-$freq$. This considerable difference can be accounted for the fact that $mbe$ can represent both harmonic and non-harmonic structure of a bird call, whereas $dom$-$freq$ in itself cannot completely justify for the non-harmonic structure. Thus we only report and analyze the results of $mbe$ individually and along with $dom$-$freq$ in the rest of the paper.

Initially, a study was carried out to extract features in smaller frequency bands motivated from the fact that the fundamental frequency of bird calls are in the range of 3-5 kHz~\cite{Fagerlund2004}. The $mbe$ and $dom$-$freq$ features were extracted in the extended band of 3-8 kHz to accommodate the higher order harmonics along with the fundamental frequency. The CBRNN with the band-limited features achieved a best AUC score of 89\% on the development set. Particularly, the number of false positives (FP) had increased, i.e. a number of recordings were wrongly flagged to have a bird call. This shows that in comparison to using band-limited features, the network is learning useful information of bird call being absent from the full band features.

\section{Results and Discussion}
\label{ssec:results}
% The CBRNN network and training configurations derived from the development set were used for the challenge submission. The CBRNN's were trained on the three CV on the development set as described in section \ref{ssec:data}. The evaluation was done on unseen data from different recording conditions. 
The average validation scores for the challenge submission set and their corresponding unseen test data scores for different dropout rates are presented in Table \ref{Table:2}. For the results without data augmentation, we see that across the feature classes and the dropout rates, the validation scores are comparable ($\approx95~\%$) and the test scores are seen to vary about 3.5\% across the features (highest of 87.2~\% and lowest of 83.7~\%).

To obtain a general insight on the significance of this 3.5\% we went through the results of the validation data. We thresholded the posterior probability of final maxout layer using a value of 0.5, i.e. a posterior probability higher than 0.5 signified that a bird call was present and otherwise absent. Among the 3138 validation recordings, there were 377 recordings classified wrongly. 242 of these were FP according to the ground truth. Since listening to all the wrongly classified recordings was not practical, we chose about the same ratio of recordings randomly for our listening test i.e. 70 FP and 30 false negatives (FN) recordings. By manually examining the audio files (i.e. we listened to the 70 recordings), we found that 37 of the 70 FP recordings had noticeable bird audio activity. Similarly, 7 of the 30 FN recordings had no bird activity. In total 42 of 100 (70 FP + 30 FN) recordings tested had wrong labels. Errors are obvious in any kind of manual annotation, and the classification method has to be robust to these. In the present scenario, the author is not sure how to correlate the annotation errors finding with algorithm performance comparison, and hence just presents it as an observation.

The results of the data augmentation and domain adaptation are presented in Table \ref{Table:2}. The general observation is that the proposed domain adaptation (test mixing) gives consistently better performance than the data augmentation method (blocks mixing). Another observation on how different the test data is with respect to training data can be noticed from the validation scores of test mixing. We see that they are consistently smaller than the validation scores without domain adaptation. In addition to these, the combination of both blocks and test mixing together was tried and was seen to perform poorly in comparison to no data augmentation on test data. It achieved an AUC of 80.3\% with 0.5 dropout.

\subsection{BAD challenge results}
The proposed method fared in the top performing submissions of the BAD challenge~\cite{badResults2016}. Apart from our method, there were five other submissions~\cite{Bulbul2016,MarioElias2016,Elias2016,Topel2016,Cakir2017} which stood out from the rest of the submissions and achieved an AUC score in the 88.0-88.7\% range. All these submissions used CNNs as part of their classifier, spectrogram features, and an ensemble of networks for the final submission. Four of them used time and frequency shifting for data augmentation~\cite{Bulbul2016,MarioElias2016,Elias2016,Topel2016}. Three of them performed a preprocessing step of noise reduction on the input data~\cite{Bulbul2016,Elias2016,Topel2016}. Two of them~\cite{Bulbul2016,Topel2016} mixed test data classified with high confidence to the training data for domain adaptation. The smallest network configuration among these~\cite{Topel2016} had approximately 328,000 parameters, in comparison to this our proposed method had 120 times fewer parameters. 

Our proposed data augmentation and domain adaptation methods were unique among the submissions. In terms of data augmentation, our proposal of blocks mixing did not give any advantage over not using it. While the domain adaptation method of test mixing was seen to be helpful. Finally, with respect to features, ours was the only submission which experimented with a harmonic specific feature ($dom$-$freq$).

\section{Conclusion} 
A stacked convolutional and bidirectional recurrent neural network architecture (CBRNN) was proposed for bird audio detection task. Two kinds of features and their combination were studied and the best result on test data was achieved using the log mel-band energy feature. The proposed novel domain adaptation was shown to consistently perform better than having no adaptation. The data augmentation method studied was not helpful and gave comparable results as without augmentation. The proposed method achieved an area under curve measure of 88.1\% on the unseen evaluation data, and 95.5\% on the development data.

\section*{Acknowledgment}
The research leading to these results has received funding from the European Research Council under the European Union’s H2020 Framework Programme through ERC Grant Agreement 637422 EVERYSOUND. Part of the computations leading to these results were performed on a TITAN-X GPU donated by NVIDIA. The authors also wish to acknowledge CSC-IT Center for Science, Finland, for computational resources.

\bibliographystyle{IEEEtran}
\bibliography{strings,refs}

\begin{thebibliography}{10}
\providecommand{\url}[1]{#1}
\csname url@samestyle\endcsname
\providecommand{\newblock}{\relax}
\providecommand{\bibinfo}[2]{#2}
\providecommand{\BIBentrySTDinterwordspacing}{\spaceskip=0pt\relax}
\providecommand{\BIBentryALTinterwordstretchfactor}{4}
\providecommand{\BIBentryALTinterwordspacing}{\spaceskip=\fontdimen2\font plus
\BIBentryALTinterwordstretchfactor\fontdimen3\font minus
  \fontdimen4\font\relax}
\providecommand{\BIBforeignlanguage}[2]{{%
\expandafter\ifx\csname l@#1\endcsname\relax
\typeout{** WARNING: IEEEtran.bst: No hyphenation pattern has been}%
\typeout{** loaded for the language `#1'. Using the pattern for}%
\typeout{** the default language instead.}%
\else
\language=\csname l@#1\endcsname
\fi
#2}}
\providecommand{\BIBdecl}{\relax}
\BIBdecl

\bibitem{Marques2012}
T.~A. Marques \emph{et~al.}, ``Estimating animal population density using
  passive acoustics,'' in \emph{Biological reviews of the Cambridge
  Philosophical Society}, vol.~88, no.~2, 2012, pp. 287--309.

\bibitem{Furnas2014}
B.~J. Furnas and R.~L. Callas, ``Using automated recorders and occupancy models
  to monitor common forest birds across a large geographic region,'' in
  \emph{Journal of Wildlife Management}, vol.~79, no.~2, 2014, p. 325–337.

\bibitem{Graciarena2011}
M.~Graciarena, M.~Delplanche, E.~Shriberg, and A.~Stolcke, ``Bird species
  recognition combining acoustic and sequence modeling,'' in \emph{IEEE
  International Conference on Acoustics, Speech and Signal Processing
  (ICASSP)}, 2011.

\bibitem{Stowell2014}
D.~Stowell and M.~D. Plumbley, ``Automatic large-scale classification of bird
  sounds is strongly improved by unsupervised feature learning,'' in
  \emph{PeerJ}, vol.~2, no. e488, 2014.

\bibitem{bad2016}
D.~Stowell, M.~Wood, Y.~Stylianou, and H.~Glotin, ``Bird detection in audio: A
  survey and a challenge,'' in \emph{IEEE Workshop on Machine Learning for
  Signal Processing (MLSP)}, 2016.

\bibitem{emre_TASLP2016}
E.~\c{C}ak{\i}r, G.~Parascandolo, T.~Heittola, H.~Huttunen, and T.~Virtanen,
  ``Convolutional recurrent neural networks for polyphonic sound event
  detection,'' in \emph{IEEE/ACM TASLP Special Issue on Sound Scene and Event
  Analysis}, 2017, accepted for publication.

\bibitem{sainath2015}
T.~N. Sainath, O.~Vinyals, A.~Senior, and H.~Sak, ``Convolutional, long
  short-term memory, fully connected deep neural networks,'' in \emph{IEEE
  International Conference on Acoustics, Speech and Signal Processing
  (ICASSP)}, 2015.

\bibitem{Choi2016}
K.~Choi, G.~Fazekas, M.~Sandler, and K.~Cho, ``Convolutional recurrent neural
  networks for music classification,'' in \emph{IEEE International Conference
  on Acoustics, Speech and Signal Processing (ICASSP)}, 2017.

\bibitem{Adavanne2017}
S.~Adavanne, P.~Pertil{\"a}, and T.~Virtanen, ``Sound event detection using
  spatial features and convolutional recurrent neural network,'' in \emph{IEEE
  International Conference on Acoustics, Speech and Signal Processing
  (ICASSP)}, 2017.

\bibitem{Nowicki1997}
S.~Nowicki, ``Bird acoustics,'' in \emph{Encyclopedia of Acoustics}, M.~J.
  Crocker, Ed.\hskip 1em plus 0.5em minus 0.4em\relax Hoboken, NJ, USA: John
  Wiley \& Sons, Inc., 2007, vol.~4, ch. 150, pp. 1813--1817.

\bibitem{jos}
\BIBentryALTinterwordspacing
J.~O. Smith, \emph{Sinusoidal Peak Interpolation, in Spectral Audio Signal
  Processing}, accessed 16.01.2017. [Online]. Available:
  \url{https://ccrma.stanford.edu/~jos/sasp/Sinusoidal_Peak_Interpolation.htm}
\BIBentrySTDinterwordspacing

\bibitem{librosa}
\BIBentryALTinterwordspacing
B.~McFee \emph{et~al.}, ``librosa v0.4.1.'' [Online]. Available:
  \url{http://dx.doi.org/10.5281/zenodo.32193}
\BIBentrySTDinterwordspacing

\bibitem{Goodfellow2013}
I.~J. Goodfellow, D.~Warde-Farley, M.~Mirza, A.~Courville, and Y.~Bengio,
  ``Maxout networks,'' in \emph{International Conference on Machine Learning
  (ICML)}, 2013.

\bibitem{batchNorm}
S.~Ioffe and C.~Szegedy, ``Batch normalization: Accelerating deep network
  training by reducing internal covariate shift,'' \emph{CoRR}, vol.
  abs/1502.03167, 2015.

\bibitem{adamKeras}
D.~Kingma and J.~Ba, ``Adam: A method for stochastic optimization,'' in
  \emph{arXiv:1412.6980 [cs.LG]}, 2014.

\bibitem{Dropout}
N.~Srivastava, G.~Hinton, A.~Krizhevsky, I.~Sutskever, and R.~Salakhutdinov,
  ``Dropout: A simple way to prevent neural networks from overfitting,'' in
  \emph{Journal of Machine Learning Research (JMLR)}, 2014.

\bibitem{chollet2015keras}
F.~Chollet, ``Keras v1.1.2,'' \url{https://github.com/fchollet/keras}, 2015.

\bibitem{theano}
\BIBentryALTinterwordspacing
{Theano Development Team}, ``{Theano: A {Python} framework for fast computation
  of mathematical expressions},'' \emph{arXiv e-prints}, vol. abs/1605.02688,
  May 2016. [Online]. Available: \url{http://arxiv.org/abs/1605.02688}
\BIBentrySTDinterwordspacing

\bibitem{giam2016}
G.~Parascandolo, H.~Huttunen, and T.~Virtanen, ``Recurrent neural networks for
  polyphonic sound event detection in real life recordings,'' in \emph{IEEE
  International Conference on Acoustics, Speech and Signal Processing
  (ICASSP)}, 2016.

\bibitem{Cakir2017}
E.~\c{C}ak{\i}r, S.~Adavanne, G.~Parascandolo, K.~Drossos, and T.~Virtanen,
  ``Convolutional recurrent neural networks for bird audio detection,'' in
  \emph{European Signal Processing Conference (EUSIPCO)}, 2017, submitted.

\bibitem{auc}
A.~P. Bradley, ``The use of the area under the roc curve in the evaluation of
  machine learning algorithms,'' \emph{Pattern Recognition}, vol.~30, no.~7,
  pp. 1145 -- 1159, 1997.

\bibitem{Fagerlund2004}
S.~Fagerlund, ``Acoustics and physical models of bird sounds,'' in
  \emph{Seminar in acoustics, HUT, Laboratory of Acoustics and Audio Signal
  Processing}, 2004.

\bibitem{badResults2016}
\BIBentryALTinterwordspacing
``Bird audio detection challenge,'' 2016. [Online]. Available:
  \url{http://machine-listening.eecs.qmul.ac.uk/bird-audio-detection-challenge-results/}
\BIBentrySTDinterwordspacing

\bibitem{Bulbul2016}
\BIBentryALTinterwordspacing
T.~Grill, ``Source code: Bulbul,'' 2016. [Online]. Available:
  \url{https://jobim.ofai.at/gitlab/gr/bird_audio_detection_challenge_2017/}
\BIBentrySTDinterwordspacing

\bibitem{MarioElias2016}
\BIBentryALTinterwordspacing
M.~Lasseck and E.~Sprengel, ``Technical report: {MarioElias},'' accessed
  10.02.2017. [Online]. Available:
  \url{http://machine-listening.eecs.qmul.ac.uk/wp-content/uploads/sites/26/2017/01/MarioElias.pdf}
\BIBentrySTDinterwordspacing

\bibitem{Elias2016}
\BIBentryALTinterwordspacing
E.~Sprengel, ``Technical report: Elias,'' accessed 10.02.2017. [Online].
  Available: \url{http://ceur-ws.org/Vol-1609/16090547.pdf}
\BIBentrySTDinterwordspacing

\bibitem{Topel2016}
\BIBentryALTinterwordspacing
T.~Pellegrini, ``Source code: Topel,'' accessed 10.02.2017. [Online].
  Available: \url{https://github.com/topel/bird_audio_detection_challenge}
\BIBentrySTDinterwordspacing

\end{thebibliography}
\end{document}